\documentclass{PoS}

\usepackage[binary-units=true]{siunitx}
\DeclareSIUnit\century{century}
\DeclareSIUnit\year{yr}
\DeclareSIUnit\month{month}
\DeclareSIUnit\week{week}
\DeclareSIUnit\day{day}
\DeclareSIUnit\particle{particle}
\DeclareSIUnit\photons{ph}
\DeclareSIUnit\photoelectrons{pe}
\DeclareSIUnit\pixel{pix}
\DeclareSIUnit\GTU{GTU}
\DeclareSIUnit\parsec{pc}
\DeclareSIUnit\lightyear{ly}

\usepackage{enumitem}
\usepackage{subcaption}

\title{Machine Learning Approach for Air Shower Recognition in EUSO-SPB Data}

\ShortTitle{Machine Learning Approach for Air Shower Recognition in EUSO-SPB Data}

\author{Michal Vr\'{a}bel$^a$, J\'{a}n Gen\v{c}i$^{a}$, Pavol Bobik$^{b}$, \speaker{Francesca Bisconti}$^{,c}$, for~the~JEM-EUSO~Collaboration\footnote{for collaboration list see PoS(ICRC2019)1177}\\
	\llap{$^a$} Faculty of Electrical Engineering and Informatics, Technical University of Ko\v{s}ice, Slovakia\\
	\llap{$^b$} Institute of Experimental Physics, Slovak Academy of Sciences, 
	Slovakia\\
	\llap{$^c$} INFN, Section of Turin, Italy\\
	E-mail: \email{michal.vrabel@tuke.sk}, \email{jan.genci@tuke.sk}, \email{bobik@saske.sk}, \email{fbiscont@to.infn.it}}

\abstract{The main goal of The Extreme Universe Space Observatory on a Super Pressure Balloon (EUSO-SPB1) was to observe from above extensive air showers caused by ultra-high energy cosmic rays. 
EUSO-SPB1 uses a fluorescence detector that observes the atmosphere in a nadir observation mode from a near space altitude. 
During the 12-day flight, an onboard first level trigger detected more than \num{175000} candidate events. 
This paper presents an approach to recognize air showers in this dataset. 
The approach uses a feature extraction method to create a simpler representation of an event
and then it uses established machine learning techniques to classify data into at least two classes - shower and noise. 
The machine learning models are trained on a set of air shower simulations put on top of the
background observed during the flight and a set of events from the flight. 
We present the efficiency of the method on datasets of simulated events. 
The flight data events are also used in unsupervised learning methods to identify groups of events with similar features. 
The presented methods allow us to shorten the candidate events list and, thanks to the groups of similar events identified by the unsupervised methods, the classification of the triggered events is made simpler.}

\FullConference{36th International Cosmic Ray Conference -ICRC2019-\\
		July 24th - August 1st, 2019\\
		Madison, WI, U.S.A.}

\begin{document}
	\setcounter{page}{2}

\section{Introduction}

EUSO-SPB1 experiment is a path-finder mission within the Joined Experiment Missions for the Extreme Universe Space Observatory (JEM-EUSO).
EUSO-SPB1 uses a fluorescence detector that observes the atmosphere in a nadir observation mode from a near
space altitude. A full description of the mission is given in the paper of Wiencke  \cite{Wiencke2017EusoSpb1MissionAndScienceIcrc}.  
The detector consists of 2 Fresnel lenses which focus the image on a pixel-matrix of photomultipliers capable of single photoelectron counting in temporal frames of \SI{2.5}{\micro\second}, named Gate Time Units (GTU). The detector's focal surface consists of a single Photo Detector Module (PDM) with $48 \times 48$ pixels. The module is organized in Elementary Cells (EC) with $16 \times 16$ pixels. The EC itself is a block of $2 \times 2$  Multi-Anode Photomultipliers (MA-PMT). The hardware-implemented First Level Trigger (FLT) \cite{Batisti2018PerformanceEusoSpbTrigger} operating on a level of EC provides an initial filtration of the data. The trigger is optimized towards storage and bandwidth and it is not designed for a detailed pattern recognition. Therefore, after the 12-day flight it produced more than \num{175000} candidate events. The events have been analyzed in several aspects \cite{Eser2019EusospbResults}.


Due to shortened duration of the mission the expected number of observed air shower events is one or fewer \cite{Shinozaki2019EusospbExposureAirShower}. 
The goal of this work is to develop an approach that would reduce the candidate events list into a smaller subset that is eventually easier to review manually.
A similar work has also been done independently of this work by Diaz at al. \cite{Diaz2019FlightDataClassification}.

The approach described in this paper uses a feature extraction method to create a simpler representation of an event
and then it uses established machine learning techniques to classify data into two classes -
\textit{"air shower"} and \textit{"noise"}. 
Below we present the efficiency of the method on datasets of simulated events. 
We also discuss the usage of the flight data events in unsupervised learning methods to identify clusters of events with similar features.

\section{Model training and testing dataset}

The dataset used for training and testing of machine learning models consisted of  simulated EUSO-SPB1 data with air shower events (positive samples) and various noise samples from real EUSO-SPB1 acquisition data (negative samples). The dataset had two main categories of the negative samples: background noise frame sequences and a small subset of labeled triggered events from the acquisition data, which were known not to be the air shower events.

The samples handled by the models were created by a procedure which, if necessary, transforms input data into EUSO-SPB-like packets. 
An acquisition data packet consists of \num{128} frames, where at frame \num{40} the hardware trigger activation happens, thus there are background noise frames available from every acquisition packet.
Simulated packets were created by selecting a sequence of first \num{32} frames from an acquisition data packet and repeating this sequence for the whole 128 GTU of the packet. The repetition was not expected to cause large biases, but other approaches should also be explored in the future.
All data were processed by a software implementation of the FLT, and the trigger data were used for temporal segmentation of a packet. This aimed to extract only a part of the packet, in this work called \textit{"the event frame sequence"}, which was expected to contain "interesting" data - ideally persistent presence of increased photo-electron counts. 
Note that photo-electron counts of the flight acquisition data considered in this analysis have been normalized by the application of the flat frame map created prior to the EUSO-SPB1 flight \cite{Eser2017EusoSpbPreflightCalibration}.

\subsection{Positive samples}

The simulation of the air shower events was done by the simulation module of the software framework ESAF \cite{Berat2010221}. The simulated primary particle energies ranged from $10^{17.6} \si{\electronvolt}$ to $10^{19} \si{\electronvolt}$ in constant steps on a logarithmic scale. Each energy step had a \num{0.05} higher exponent value than the previous one. The simulated balloon altitudes ranged from \num{18} to \SI{33}{\kilo\meter}. 
Events were simulated uniformly arriving from all directions. However, note that in a resulting training dataset, the distributions were not uniform due to applied selections based on visibility of a simulated air shower. 
The simulated events had to contain recognizable air shower tracks, and the simulated event frame sequences had to contain at least \num{2} frames, where a maximum signal pixel value was greater than a maximum background noise pixel value.  
Applying both rules on the available data processed by the feature extraction algorithm, the total number of events available for analysis was \num{34521}.

\subsection{Negative samples}

The main portion of the negative samples in the dataset were the event frame sequences triggered by the software FLT at least \num{20} frames before or after frame \num{42}. These event frame sequences are called \textit{"unlabeled noise events"}. In the whole dataset, the number of such event frame sequences was \num{56229}.
Another significant part of the dataset were the event frame sequences triggered by the software FLT less than \num{20} frames before or after frame \num{42} and classified by a manual review. 
Such event frame sequences are referred to as \textit{"labeled noise events"}.
The main difference of these events from the \textit{unlabeld noise events} was that the \textit{labeled flight events} were typically activations of the trigger due to high intensity peaks in the photo-electron counts. 
For the purpose of the air shower detection, the only important information was that these events are not air showers. 
However, several classes were identified on an intuitive basis. The most typically assigned classes of such events were the following: \textit{"a single pixel"}, \textit{"the top-left elementary cell"}, \textit{"a suddenly increased background intensity"}, \textit{"a blob"}, \textit{"a large blob"}, \textit{"a bright blob"}, \textit{"the noise"}, and \textit{"a single GTU track"}. 


\section{Feature extraction based approach}

The feature extraction approach classified the event frame sequences in two general steps: feature extraction and classifier training. Following the temporal segmentation of the event frame sequence using data from the software FLT, this approach applied a feature extraction procedure which produced a set of features describing various properties of the sequence. Afterwards, the results were stored in a database table prepared for the analysis.
The air showers or other types of events could then be searched by SQL queries specifying feature thresholds. However, such manual classification experiments did not yield satisfactory results. Therefore, it was decided to use a machine learning approach to perform a binary classification. The machine learning algorithm would, based on the training dataset, in a sense formulate such conditions.

\subsection{Feature extraction procedure}\label{sec:FeatureExtractionProcedure}

The feature extraction procedure uses X-Y, GTU-X, GTU-Y projections of an event and processes them by a Hough transform with the normal parametrization \cite{duda-hough}.  
%
A sequence of frames can be viewed as a three dimensional matrix. 
The projection is created by selecting a maximum value in a dimension not visible in that projection - it is the time in X-Y projection (a maximum value of a pixel within a sequence of frames), a column of PDM pixels in GTU-X, and a row of PDM pixels in GTU-Y projection.  An inclined line with an inclination different from \ang{0} and \ang{90} indicates movement of a bright spot, \ang{0} line indicates no movement, and \ang{90} line indicates pixel value increase appearing only in a single frame (e.g. a track appearing only in a single frame).

The transform creates a parameter space (Hough space) which for each line associates line parameters (angle of the line's normal and the line's distance from the origin) with the sum of values of pixels crossing this line in the image. 
Important parts of the procedure are also the data preparation before the transform and the method of determining an optimal line orientation from the parameter space.
Several variants of X-Y, GTU-X, GTU-Y projections are created by selecting only triggered pixels or pixels above a threshold. Several threshold levels are considered: pixels with values above mean nonzero pixel value + $\{0, 1, 2\} \times \sigma$, and pixels with values above threshold selected by the Yen method \cite{Yen1995Thresholding}. A Hough space is constructed for each of those projections. Then the space is thresholded in several levels (\SI{75}{\percent}, \SI{85}{\percent}, \SI{90}{\percent}, \SI{95}{\percent} of a peak value), and it is analyzed afterwards.

Some examples of the extracted features include the following: 
the number of frames in an event frame sequence;
the maximum and minimum pixel values;
the properties of pixel clusters in the projections; 
Hough space properties - orientation and location of the most significant line, number of clusters in the parameter space, maximum and minimum cluster dimensions in the space (width, height, area), dimensions of a cluster with the maximum value, dimensions of a cluster with the maximum total sum of pixel values.

\subsection{Classifier training}

The positive and negative datasets were joined, balanced, shuffled and split into training and testing sets in \num{60}:\num{40} ratio. 
The training set samples were also weighted such that the total weights are equal between the labeled noise samples and the unlabeled noise samples.
Each sample was initially represented by a set of more than \num{1000} features. Recursive feature elimination \cite{Guyon2002RFE} (RFE) was used to select an optimal set of features to train the extremely randomized trees classifier \cite{Geurts2006ExtraTrees}. 
Most of the selected features describe the width of the cluster in a Hough space.
Unexpectedly, sets of \num{150}-\num{250} features in any classification experiment do not include orientation of the line in the time-wise projections of the air shower. Using the dataset described above, the most important feature, out of \num{153}, is the width of a cluster in a Hough space from the X-Y projection of the event frame sequence.
The projection was thresholded by selecting pixels with intensity above the mean value, and the Hough space was thresholded by selecting pixels above \SI{75}{\percent} of maximum value in this space.
The classifier trained on this feature alone achieved \SI{79}{\percent} overall accuracy.


\subsection{The classifier accuracy on the test data}

\begin{figure}[t]
	\centering
	\begin{subfigure}[b]{0.515\textwidth}
		\centering
		\includegraphics[width=0.95\textwidth]{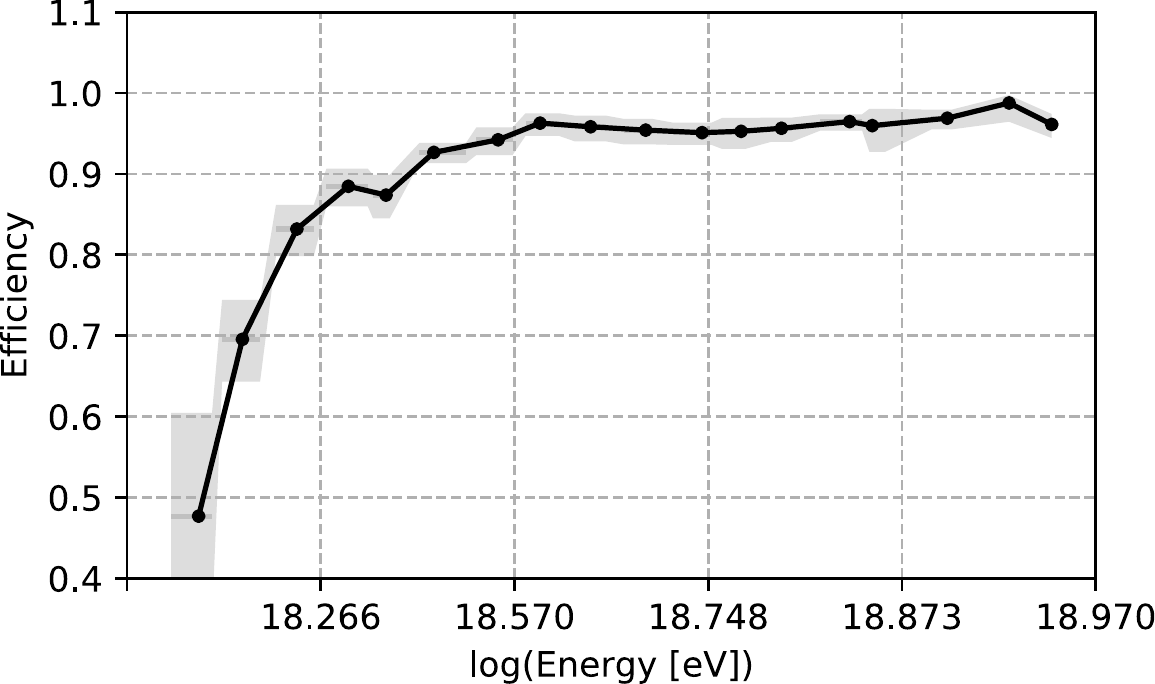}
		\caption{Efficiency as a function of the energy on the test set.}
		\label{fig:test_set_sensitivity_function_of_energy_linear_20steps}
	\end{subfigure}
	~
	\begin{subfigure}[b]{0.465\textwidth}
		\centering
		\includegraphics[width=1\textwidth]{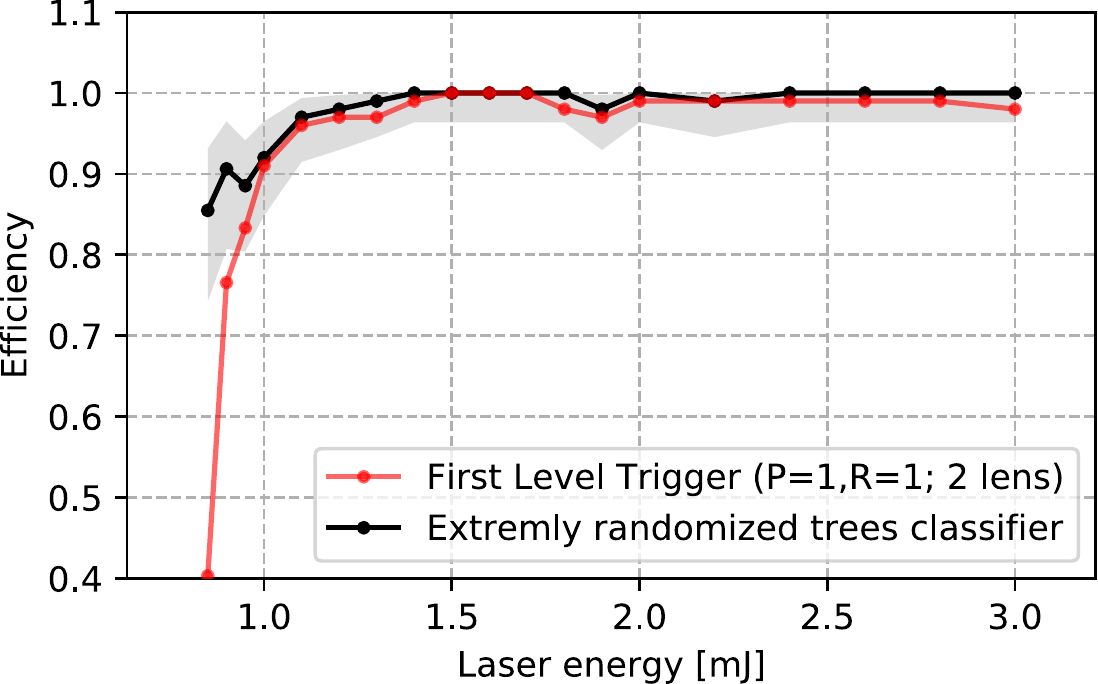}
		\caption{Efficiency as a function of the laser energy.}
		\label{fig:efficiency_plot_acq_group_041016-gls-45degaway_eusobal_classification_model_comparison_max_err_0_20}
	\end{subfigure}
	
	\caption{Efficiency of the classifier.}\label{fig:FlightDataFlatFieldingExample}
\end{figure}

Figure \ref{fig:test_set_sensitivity_function_of_energy_linear_20steps}
shows efficiency (ratio correctly classified to all simulated event frame sequences in the test set)  of 
the 
classifier 
described above
as a function of energy of a simulated air shower event. 
The grey areas on the plot show binomial proportion confidence intervals (Clopper-Pearson intervals) at \SI{95}{\percent} confidence level. 
The areas have a box-like shape because the width indicates range of the attribute values grouped under a single point. Figure \ref{fig:test_set_sensitivity_function_of_energy_linear_20steps} shows that the efficiency increases with energy of the primary particle - the plot can be approximated by the exponential function $\epsilon_{\textrm{cls}}(E) = -0.927 \exp (-1.123 \times 10^{-18} E) + 0.96$, where $E$ is the primary particle energy in \si{\electronvolt}. Furthermore, the efficiency decreases with increasing distance of a shower maximum from the center of the FoV, and the efficiency slightly increases with the zenith angle. However, note that the plot shows results on the \textit{visible tracks} dataset, if those rules were not applied, inclusion of the high-zenith angle events that are passing through an edge of the PDM might skew the results.

Overall non-balanced cross-validated accuracy of the presented classifier is approximately \SI{95}{\percent}, overall sensitivity (efficiency) is \SI{93}{\percent}, overall specificity is \SI{98}{\percent}, but specificity on the labeled noise data of the test set is approximately \SI{92}{\percent}. Based on a comparison with several previous classification experiments with a different labeled noise sample count (not shown here), the sensitivity seems to decrease as more air-shower-like noise samples are included in a training set.

\section{Air shower recognition in the dataset of laser shots}

An independent evaluation of the classifier performance was done by classifying the data from EUSO-SPB calibration and testing in 2016 \cite{Eser2017EusoSpbPreflightCalibration, Bertaina2017PerformanceEusoSpbTriggerIcrc}. The goal to this exercise was to make sure that the classifier had not missed valid air shower events passed by the FLT. Figure \ref{fig:efficiency_plot_acq_group_041016-gls-45degaway_eusobal_classification_model_comparison_max_err_0_20} represents the same trigger data as the figure 3 in the paper of Bayer et. al \cite{Bertaina2017PerformanceEusoSpbTriggerIcrc}. However, note that these are long tracks going through the middle of the focal surface. Unlike the flight data, this dataset was not normalized by the flat field map, and this might have contributed to the high number of the event frame sequences identified by the software implementation of FLT - approximately \num{4.18} times more than the actual number of packets in the acquisition run. The analysis have also displayed a weakness of the current version of the classifier - an actual number of event frame sequences classified as an air shower was about \num{1.2} times higher than number of event frame sequences which have overlap with frame \num{40}.

\newpage
\section{Air shower recognition in the EUSO-SPB flight data}

\begin{figure}[t]
	\centering
	\begin{subfigure}[b]{0.31\textwidth}
		\includegraphics[width=1.0\linewidth]{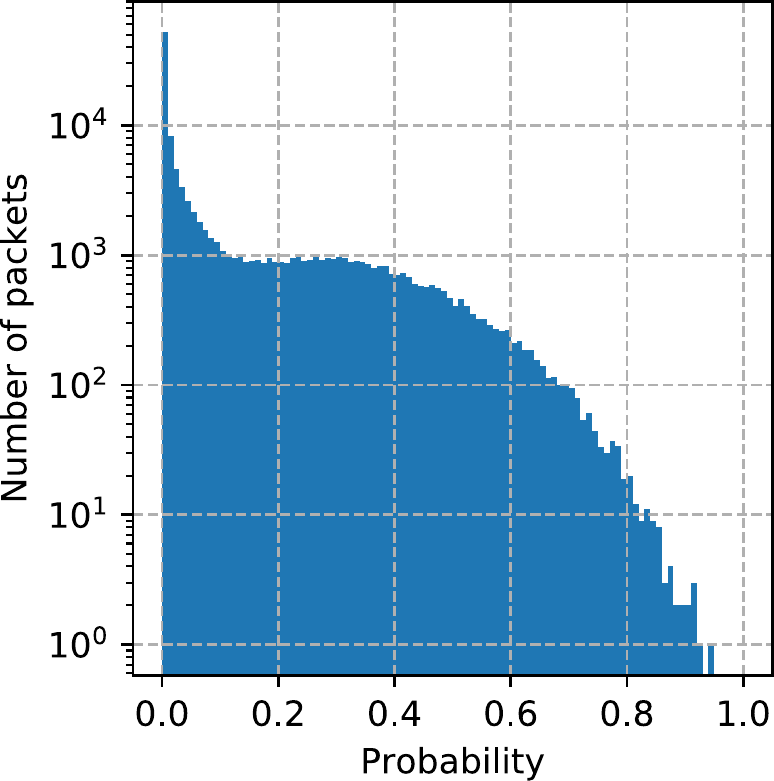}
		\caption{Probability distribution.}
		\label{fig:flight_df_gtu_36_45_pack_nonan_extra_trees_cls_on_train_rfecv_est_dropna_proba_distribution_vertical}
	\end{subfigure}
	~~~~~
	\begin{subfigure}[b]{0.51\textwidth}
		\includegraphics[width=1\linewidth]{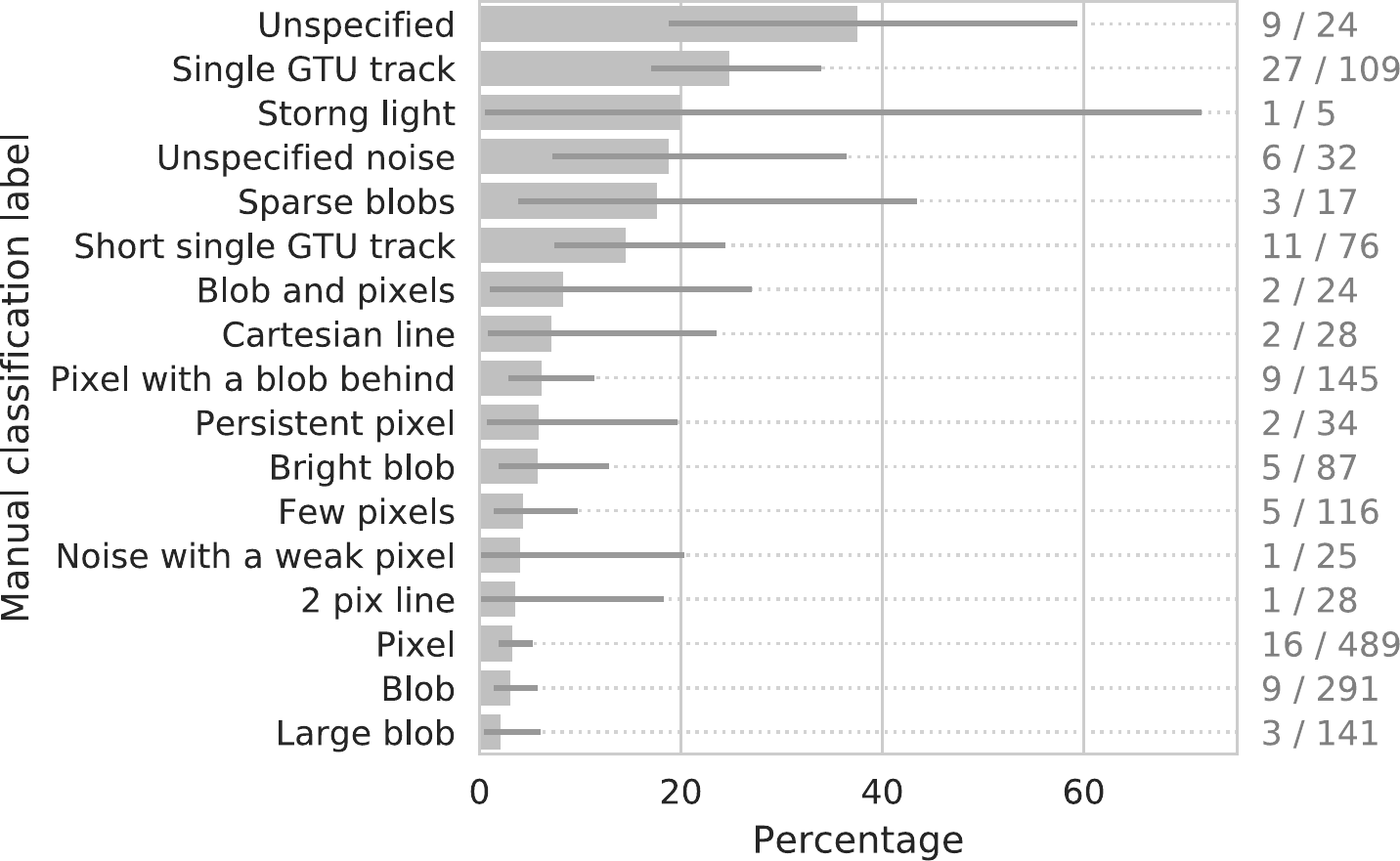}
		\caption{Percentages of selected classified events.}
		\label{fig:flight_data_manual_classification_class_label_normalized_distribution_with_unclassified}
	\end{subfigure}
	
	\caption{Classification of the EUSO-SPB flight data - samples classified as an air shower.}\label{fig:flight_data_manual_classification}
\end{figure}

The event frame sequences selected for the classification were required to contain at least \num{6} active elementary cells and the software FLT activation had to happen between frames \num{36} and \num{45}. If there were several independently identified frame sequences in this range, the sequence closer to the frame \num{40} was selected. This cut yielded \num{118141} event frame sequences (\SI{80}{\percent} of packets passed through the software trigger). After classifying the data using the extremely randomized trees classifier, \num{5450} events (\SI{4.6}{\percent}) had higher probability than \SI{50}{\percent} of being an air shower. The number of these entries seems to depend on the labeled noise entries count in the training set - in this case it was \num{2619}. A model trained without such data selected almost \SI{50}{\percent}, and comparable models, trained on a dataset with \num{1800} of such entries, selected around \SI{10}{\percent} of the classified entries. 

Figure \ref{fig:flight_df_gtu_36_45_pack_nonan_extra_trees_cls_on_train_rfecv_est_dropna_proba_distribution_vertical} shows the distribution of probabilities in the classified flight data: With the increasing probability, the number of packets decreases. There are only \num{87} packets with \SI{80}{\percent} probability and only \num{7} packets with \SI{90}{\percent} probability. Just for a comparison, all of the laser shots considered in the figure \ref{fig:efficiency_plot_acq_group_041016-gls-45degaway_eusobal_classification_model_comparison_max_err_0_20} have been classified as an air shower with higher than \SI{80}{\percent} probability.
A manual review of a subset of the samples, classified as an air shower in an order from entries with the highest probability, did not find any actual air shower.

Figure \ref{fig:flight_data_manual_classification_class_label_normalized_distribution_with_unclassified} shows the fraction of incorrectly classified to all labeled noise event frame sequences in the flight data. The plot indicates how successfully the classifier rejected false positive events in the labeled flight data subset. Error bars show binomial proportion confidence interval at \SI{95}{\percent} confidence level. The interval was calculated via the Clopper-Pearson method considering all class samples count and the selected count. These numbers are shown on the right side of the plot. Note that in case of the flight data classification, some entries were in the classifier's training set.

In this case, almost \SI{40}{\percent} of events classified \textit{"unspecified"} have been selected. Out of an abundance of caution, these are types of events that have not been included in the training set. Although, these are not really considered to be the strong air shower candidates, the events were considered to share some similarity with an air shower pattern. Another typical false positives are \textit{"single GTU tracks"}. In all realized classification experiments so far, this class has been one of most typical false positives. An increase in a number of the events in the training set so far decreased the number of false positive classifications. 


\section{An attempt to understand the data via T-SNE}

T-distributed stochastic neighbor embedding (T-SNE) \cite{Maaten2008TSNE} is a non-linear dimensionality reduction  technique. The technique iteratively optimizes point positions in a low-dimensional space (embedding) to ensure that they are similar if these points are similar in the original high-dimensional space. In this application, the points represent event frame sequences in the dataset.
The technique provides a tool to evaluate whether the selected features capture the differences between the air showers and the noise. Although not shown here, the visualization of the T-SNE embedding on the testing and training data showed separation between the classes for most of the samples.
On the other hand, the application of the technique on the unlabeled flight data with the small set of the labeled noise events 
helps us to better understand the dataset contents.
The approach also allows us to visualize areas where the classifier tends to make false positive detections.



\begin{figure}[t]
	\centering
	\captionsetup[subfigure]{format=hang,justification=raggedright}
	
	\begin{subfigure}[b]{0.236\textwidth}
		\centering
		\includegraphics[width=1\linewidth]{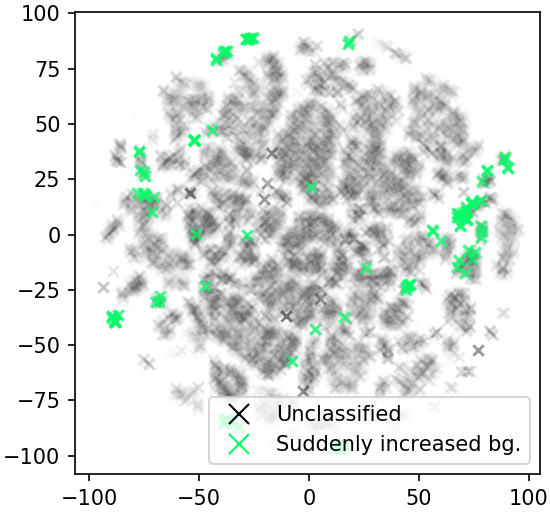}
	\caption{Suddenly increased bg. intensity events.}
	\label{fig:flight_data_gtu_36_45_pack_feat_153_tsne_dbscan_1_90_clusters_for_bg_increased_suddenly_scatter}
	\end{subfigure}
	~
	\begin{subfigure}[b]{0.236\textwidth}
		\centering
		\includegraphics[width=1\linewidth]{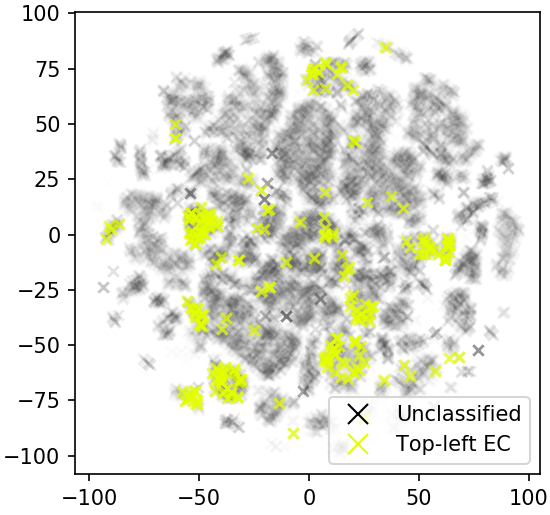}
		\caption{Top-left elementary cell problem events.}
		\label{fig:flight_data_feat_153_tsne_dbscan_1_90_clusters_for_top_left_ec_scatter}
	\end{subfigure}
	~
	\begin{subfigure}[b]{0.236\textwidth}
		\centering
		\includegraphics[width=1\linewidth]{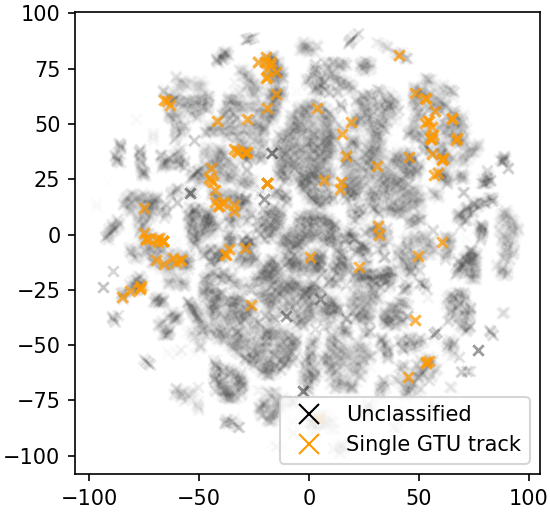}
		\caption{Single GTU track events.}
		\label{fig:flight_data_feat_153_tsne_dbscan_1_90_clusters_for_single_gtu_track_scatter}
	\end{subfigure}
	~
	\begin{subfigure}[b]{0.236\textwidth}
	\centering
	\includegraphics[width=1\linewidth]{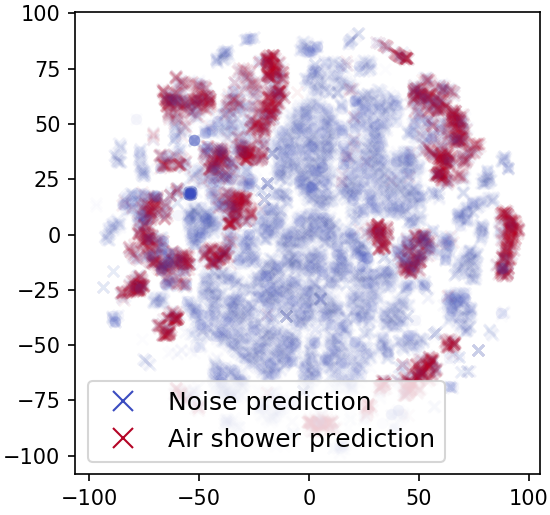}
	\caption{Probable failures of the classifier.}
	\label{fig:flight_data_feat_153_tsne_high_intensity_shower_15_inches}
	\end{subfigure}

\caption{T-SNE of the flight data using the feature set optimized for the air shower classification.}\label{fig:flight_data_feat_153_tsne}
\end{figure}

Figure \ref{fig:flight_data_feat_153_tsne_high_intensity_shower_15_inches} illustrates the application of the technique on the EUSO-SPB1 flight data. 
Note, that the cluster sizes are not well representing the number of samples in a cluster, and the actual absolute coordinates of the points are not very important.
The first three figures show locations of manually labeled entries in the embedding. 
Types of events, such as an increase of pixel intensity over the whole PDM (fig. \ref{fig:flight_data_gtu_36_45_pack_feat_153_tsne_dbscan_1_90_clusters_for_bg_increased_suddenly_scatter}) or a malfunction of top-left elementary cell (fig. \ref{fig:flight_data_feat_153_tsne_dbscan_1_90_clusters_for_top_left_ec_scatter}) are well identified and generally in the separate clusters. Single GTU track events (fig. \ref{fig:flight_data_feat_153_tsne_dbscan_1_90_clusters_for_single_gtu_track_scatter}) are not as well separated but there is an observable clustering. The last figure (fig \ref{fig:flight_data_feat_153_tsne_high_intensity_shower_15_inches}) shows locations where the classifier marked the data as an air shower. Those should be investigated in detail, and eventually put into the training set to better optimize the classifier.


\section{Conclusion and outlook}

The paper describes one approach used in search for air shower events in the dataset of EUSO-SPB1. The approach relies on the features extracted by a dedicated feature extraction program. 
The classifier has achieved \SI{95}{\percent} cross-validated balanced accuracy in detection of the air shower events in the testing dataset. Its comparison with the FLT efficiency on laser shot data have shown that the air shower events triggered by the FLT should not be rejected by the classifier. 
None of the currently applied approaches have found any air shower event in the EUSO-SPB1 flight data.

An alternative approach is to train, for instance, a neural network directly on the raw data. Such approach has also been investigated and it will be presented in future publications. 
Further development of the air shower recognition methods will continue. Eventually, the methods will be important for the upcoming EUSO-SPB2 mission presently planned for the flight in \num{2022}.



\section{Acknowledgment}

This work was partially supported by NASA grants NNX13AH54G, NNX13AH52G, French Space
Agency (CNES), Italian Space Agency through the ASI INFN agreement n. 2017-8-H.0, Italian
Ministry of Foreign Affairs and International Cooperation, the Basic Science Interdisciplinary Research Projects of RIKEN and JSPS KAKENHI Grant (22340063, 23340081, and 24244042),
Deutsches Zentrum f\"{u}r Luft und Raumfahrt, 'Helmholtz Alliance for Astroparticle Physics
HAP' funded by the Initiative and Networking Fund of Helmholtz Association (Germany), and by Slovak
Academy of Sciences MVTS JEM-EUSO as well as VEGA grant agency project 2/0132/17. We acknowledge NASA Balloon Program Office, Columbia Scientific Balloon Facility, Telescope Array
Collaboration, and Wanaka airport. The present research used resources of the National Energy
Research Scientific Computing Center (NERSC), a US Department of Energy Office of Science
User Facility operated under Contract No. DE-AC02-05CH11231.

\bibliographystyle{JHEP}
\bibliography{bibliography.bib}

\end{document}